\documentclass[sn-mathphys,Numbered]{sn-jnl}

\usepackage{graphicx}%
\usepackage{multirow}%
\usepackage{amsmath,amssymb,amsfonts}%
\usepackage{amsthm}%
\usepackage{mathrsfs}%
\usepackage[title]{appendix}%
\usepackage{xcolor}%
\usepackage{textcomp}%
\usepackage{manyfoot}%
\usepackage{booktabs}%
\usepackage{algorithm}%
\usepackage{algorithmicx}%
\usepackage{algpseudocode}%
\usepackage{listings}%

\theoremstyle{thmstyleone}%
\theoremstyle{thmstyletwo}%

\theoremstyle{thmstylethree}%

\begin{document}

\title[Article Title]{Quantitative phase imaging of opaque specimens with flexible endoscopic microscopy}


\author[1]{\fnm{Jingyi} \sur{Wang}}
\author[1]{\fnm{Wu} \sur{You}}
\author[1]{\fnm{Yuheng} \sur{Jiao}}
\author[2]{\fnm{Yanhong} \sur{Zhu}}
\author[1]{\fnm{Xiaojun} \sur{Liu}}
\author[1,3]{\fnm{Xiangqian} \sur{Jiang}}
\author*[4]{\fnm{Chenfei} \sur{Hu}}\email{chenfei\_hu@amat.com}
\author*[1,5,6]{\fnm{Wenlong} \sur{Lu}}\email{hustwenlong@mail.hust.edu.cn}

\affil[1]{\orgdiv{State Key Laboratory of Digital Manufacturing Equipment and Technology, School of Mechanical Science and Engineering}, \orgname{Huazhong University of Science and Technology}, \orgaddress{ \city{Wuhan}, \postcode{430074}, \country{PR China}}}

\affil[2]{\orgdiv{National Engineering Research Center for Nanomedicine, College of Life Science and Technology}, \orgname{Huazhong University of Science and Technology}, \orgaddress{ \city{Wuhan}, \postcode{430074}, \country{PR China}}}

\affil[3]{\orgdiv{EPSRC Future Metrology Hub}, \orgname{EPSRC Future Metrology Hub}, \orgaddress{\city{Huddersfield}, \postcode{HD1 3DH}, \country{UK}}}

\affil[4]{\orgname{Applied Materials, Inc.}, \orgaddress{ \city{Santa Clara}, \postcode{95054}, \country{USA}}}

\affil[5]{\orgdiv{Shenzhen-HUST Research Institute}, \orgname{Huazhong University of Science and Technology}, \orgaddress{ \city{Shenzhen}, \postcode{518057}, \country{PR China}}}

\affil[6]{\orgname{Optics Valley Laboratory}, \orgaddress{ \city{Wuhan}, \postcode{430074}, \country{PR China}}}


\abstract{The flexible endoscope is a minimally invasive tool in clinical settings, but most of them rely on exogenous staining for diagnosis to provide qualitative information. Here, we demonstrated a flexible endoscopic microscopy (FEM) with diffracted gradient light for quantitative phase imaging of unlabeled thick samples. Our instrument features a small form factor fiber bundle as the endoscope probe, cellular-level lateral and axial resolutions, and direct phase measurement via simple field modulation. By testing pathologic slices, thick opaque mammalian tissue ex vivo and wound healing in vivo, FEM identifies normal and tumor glandular structures, secreta, and tomographic skin layers. With the advantages of direct morphological and phase measurement, high resolution, and thin fiber tip, the label-free FEM could be an attractive tool for various clinical applications.}

\maketitle
\section{Introduction}\label{sec1}
Flexible endoscope is a significant clinical tool for diagnosis and therapeutics \cite{Mannath_2016,li2018perspective,Kurniawan_2017}, due to great surgical freedom, minimally invasion and high imaging quality. As conventional digital endoscope only provides amplitude imaging of the surface, novel techniques like confocal endoscopy, multiphoton endoscopy and optical coherence tomography endoscopy have been developed for depth sectioning recently. Multiphoton endoscopy and confocal laser endoscopy provide cell-level resolution images with specific information revealed by fluorescent tags  \cite{klioutchnikov2020threephoton,kucikas2021twophoton,loterie2015confocal,zhang2019line,wen2023single}. However, aside from intrinsic fluorescence imaging, these techniques mostly require exogenous labeling and suffer from phototoxicity and photobleaching. Optical coherence tomography provides label-free images with a deep depth of imaging, but often needs lateral scanning modules to get \textit{en face} images and provides back scattering amplitude information  \cite{li2020ultrathin,gora2017endoscopic}.

Phase is an invaluable endogenous contrast mechanism that captures optical path length shifts resulting from variations in the refractive index of a sample. Quantitative phase imaging (QPI) is a fascinating technique in biomedicine that provides cellular structural information and achieves phase mapping of biophysical parameters with nanometer-scale sensitivity \cite{park2018quantitative}. This technique has numerous biomedical applications, including live cell imaging, cancer diagnosis, hematology, and pathology \cite{park2018quantitative,kemper2008digital,merola2016tomographic,nygate2020holographic,rivenson2019phasestain,hu2022live,kim2022rapid,CACACE2020106188,https://doi.org/10.1002/cyto.a.23100}. However, most QPI techniques use trans-illumination and cannot penetrate deep into thick or opaque specimens, limiting their potential applications in clinical settings. Several reflection-mode QPI techniques with optical sectioning ability have emerged to quantify fluctuations of cells and nuclei at the nanometer scale or show potential for thick specimens imaging \cite{singh2019studyinga,choi2018reflectiona,zhou2017modelinga,kandel2019epiillumination}, but they are not directly applicable for quantitative endoscopic imaging.

Efforts have been devoted to endoscopic QPI systems, such as rigid endoscopic diffraction phase microscopy and flexible holographic endoscopes. Rigid endoscopic diffraction phase microscopy \cite{hu2018endoscopic} enables real time single shot imaging, but it is slow in operation and its rigid probe constraints its accessibility to certain tissue types, such as sinuous areas. Flexible holographic endoscopes  \cite{sun2022quantitative,badt2022realtime,choi2020fourier}achieve high resolution images with an ultra-thin lensless fiber tip. However,  these techniques require computationally expensive system calibration or phase reconstruction, which compromises temporal stability and throughput. The quantitative oblique back illumination fiber endoscope  \cite{ford2012phasegradient,doi:10.1063/1.4941208,costa2021invivo}, a speckle-free system using LED illumination, shows morphological differences between normal and tumor regions of thick brain tissue. The phase reconstruction process relies on numerical modeling of an ideal optical system and thus becomes computationally expensive.

  In this paper, we develop a flexible endoscopic microscopy (FEM) with a 2.5 mm diameter tip, which works in reflection geometry to achieve high-resolution \textit{en face} imaging of biological samples. In FEM system, an amplitude-type SLM is utilized to introduce lateral shearing and phase shifting, which enables measuring phase directly without complicated transfer function modeling and reconstruction calculation. By measuring microspheres and patterned structures on a scattering substrate, FEM demonstrates cellular level resolution in both transverse and axial directions. Furthermore, by measuring histological slides, mammalian skin tissue \textit{ex vivo}, and wounded skin tissue \textit{in vivo}, the phase images measured by FEM can clearly identify normal and tumor glandular structures, fibrous components, and different skin layers. The non-destructive tomographic investigation ability, cellular resolution, and potential diagnosis information provided by FEM might make it an interesting tool for clinical applications.
  
  \section{Results}
\subsection{FEM System Design}

\begin{figure}[h]%
\centering
\includegraphics[width=0.9\textwidth]{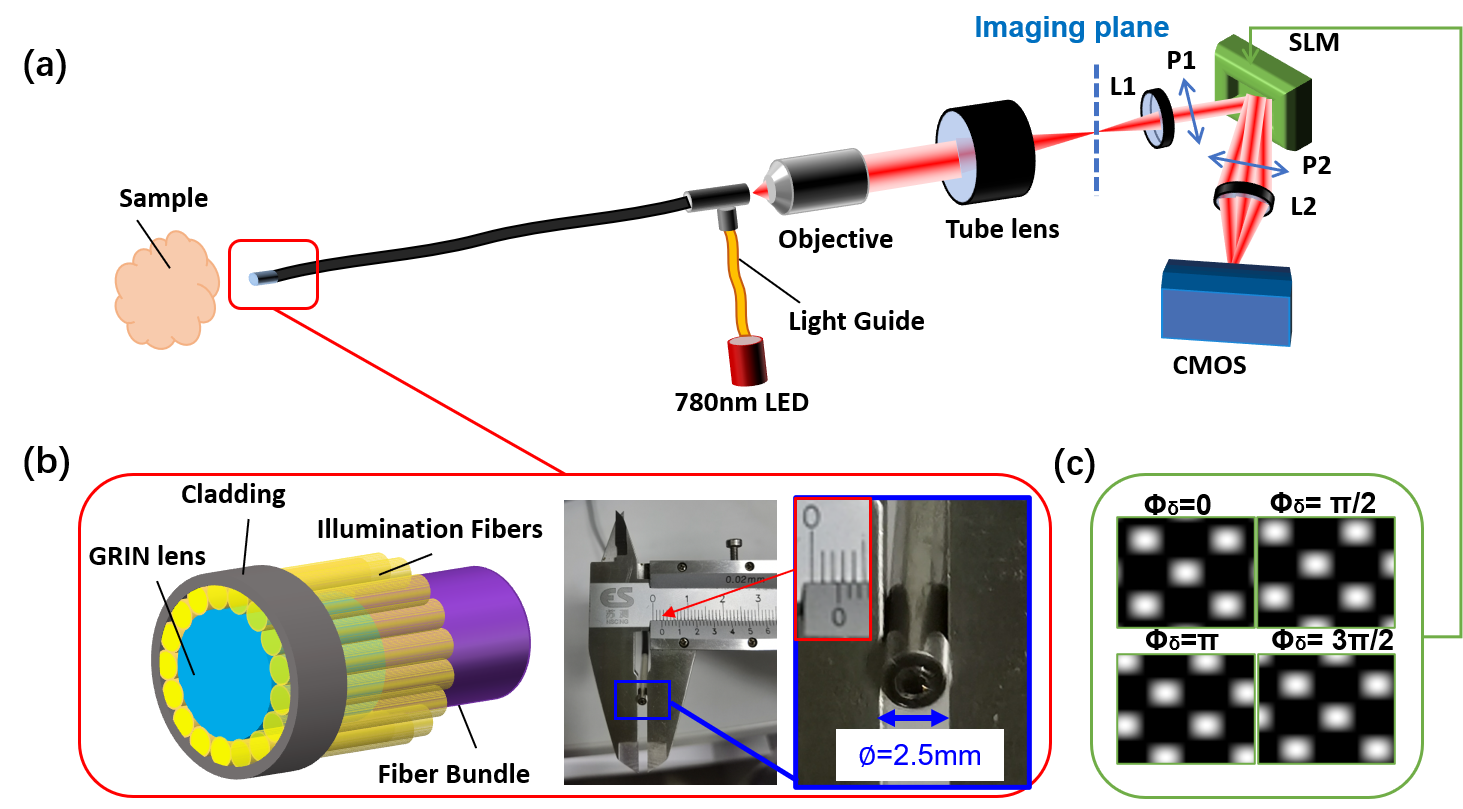}
	\caption{Schematic representation of the FEM system. (a) Optical setup: LED light is transmitted through the light guide for illumination, while the objective lens, accompanied by the tube lens, collects images that are transferred by the fiber bundle. L1 and L2 are lenses in a 4f system, and P1 and P2 are linear polarizers. The polarization of P1 is parallel to the liquid crystal axis, and perpendicular to that of P2, to help SLM achieve amplitude modulation. The detailed information is described in Methods. (b) View of the probe distal end. The illumination fibers surround the GRIN lens and fiber bundle to form ring illumination. The total diameter of the cladding protected fiber tip is 2.5 mm. (c) Four diffraction patterns are displayed on the SLM for phase shifting and wavefront shearing. }
	\label{fig:setup}
\end{figure}

The FEM system, illustrated in Fig.\ref{fig:setup} (a), consists of two key sub-systems: endoscopic optics and microscopic optics. The endoscopic optics are utilized for illumination and wavefront transmission, whereas the microscopic optics are employed for imaging and wavefront modulation. To achieve illumination, light from a LED source is coupled into the light guide and delivered to the illumination fibers. As depicted in Fig.\ref{fig:setup} (b), the illumination fibers are positioned surrounding the fiber bundle to create a ring illumination. The outer diameter of the probe distal end is approximately 2.5 mm. In the imaging geometry, light reflected from the sample is collected by a gradient refractive index (GRIN, 2.6×) lens and transmitted through the fiber bundle to the front focal plane of the objective. The GRIN lens, along with the objective (10×, NA=0.4) and tube lens, magnifies the sample field by a factor of 26. At the FEM output port, a 4-f system is followed to relay the magnified image to the camera plane. At the back focal plane of the $1^{st}$ Fourier lens, a spatial light modulator (SLM) is employed to modulate the reflective wavefront. Here, the SLM projects a chessboard grating pattern (Fig.\ref{fig:setup} (c)), and, thus, the interferograms of differentiated light can be captured by a CMOS camera \cite{mcintyre2010quantitative}.By controlling the grating period on the SLM, the shear distance can be well controlled below the diffraction limit. For each field of view, 4 interferograms are recorded, each with a unique SLM phase delay. More details regarding the system design and image acquisition are included in the Methods.

\subsection{Image formation and signal processing}

An image reconstruction method is employed to convert the recorded interferograms to phase images. The field at the image plane, \textit{U}, can be modeled as
\begin{equation}
 U(x,y)=A_{ij}(x,y)e^{i\varphi_{ij}(x,y)}\cdot\left[U_0(x,y)e^{i\varphi(x,y)}\right],
	\label{eq:eq01}
\end{equation}
where {$x,y$} are spatial coordinates in the transverse plane, {$U_{0}$} is the amplitude of the image field before entering the fiber bundle, {$A_{i j}$} and {$\varphi_{ij}$} are the amplitude and phase variation caused by the fiber bundle, and {$\varphi$} is target phase. At the Fourier plane, the wavefront of the light field is modulated by the chessboard grating pattern generated by the SLM, and the resulting field, {$U^\prime$}, can be formulated as

\begin{equation}
U^{\prime}(k_x,k_y)=U(k_x,k_y)\cdot\left[\cos\left(k_x\cdot x_0+\frac\delta2\right)\cdot\cos\left(k_y\cdot y_0+\frac\delta2\right)+DC\right],
	\label{eq:eq02}
\end{equation}

where {${k}_{x}$} and {${k}_{y}$} are the spatial frequency in the $x$ and $y$ directions, {$U(k_x,k_y)$} indicates the Fourier transform of {$U(x,y)$}, {$x_{0}$} and {$y_{0}$} are shear distances introduced by the grating that equaling {$\frac{d\cdot\lambda}{f\cdot\mathbf{M}}$}, where $d$ is the grating period, $\lambda$ is the wavelength, $f$ is the focal distance of L4, and M is the magnification of the system. {${\delta}$} is the additional phase shifts introduced by SLM, {$DC$} is the $0^{th}$ diffraction order caused by SLM. Thus, the field at the camera plane, {$U^\prime$}, takes the form

\begin{equation}
\begin{aligned}U^{^{\prime}}(x,y)&=U(x+x_0,y+y_0)e^{i\delta}+U(x-x_0,y-y_0)e^{-i\delta}+U(x+x_0,y-y_0)\\&+U(x-x_0,y+y_0)+U(x,y)\cdot DC\end{aligned},
	\label{eq:eq03}
\end{equation}

where {$U^\prime\left(x,y\right)$} is the inverse Fourier transform part of {$U^\prime\left(k_x,k_y\right)$}. The amplitude of the image fields is assumed to be uniform when the shear distance is smaller than the diffraction limit, meaning {$\left|U\left(x\pm x_0,y\pm y_0\right)\right|=\left|U\left(x,y\right)\right|$}. The gradient phase of {$U(x,y)$} can be easily calculated using 4 interferograms with distinctive phase shift 
\begin{subequations}
    \begin{gather}
  \Delta \varphi^{\prime}=\Delta \varphi_{\mathrm{ij}}(x, y)+\Delta \varphi(x, y)=\frac{\left|U^{\prime}(x, y)\right|_{\delta=3 \pi / 2}^{2}-\left|U^{\prime}(x, y)\right|_{\delta=\pi / 2}^{2}}{\left|U^{\prime}(x, y)\right|_{\delta=0}^{2}-\left|U^{\prime}(x, y)\right|_{\delta=\pi}^{2}},\label{eq:eq04a}\\
\mathrm{\Delta\varphi}\left(x,y\right)=\varphi\left(x+x_0,y+y_0\right)-\varphi\left(x,y\right),\label{eq:eq04b}\\
\mathrm{\Delta}\varphi_{ij}\left(x,y\right)=\varphi_{ij}\left(x+x_0,y+y_0\right)-\varphi_{ij}\left(x,y\right).
	\label{eq:eq04c}
 \end{gather}
\end{subequations}

Since the fiber bundle inducted phase is considered static for adjacent imaging planes with displacement within resolution level, {$\mathrm{\Delta}\varphi_{ij}$} can be directly computed from the interferograms measuring from a highly scattering medium, or planar object with uniform thickness. The phase difference of the sample from the fiber deteriorated image with {$\mathrm{\Delta\varphi}=\mathrm{\Delta}\varphi^\prime-\mathrm{\Delta}\varphi_{ij}$}, such that the sample phase {$\varphi$} is obtained via

\begin{equation}
\varphi(x,y)=FT^{-1}\left[\frac{FT\bigl(\Delta\varphi(x,y)\bigr)\cdot H^*\bigl(k_x,k_y\bigr)}{H\bigl(k_x,k_y\bigr)\cdot H^*\bigl(k_x,k_y\bigr)+\varepsilon}\right],
	\label{eq:eq05}
\end{equation}

with

\begin{equation}
H\left(k_x,k_y\right)=e^{2\pi i\left(x_0\cdot k_x+y_0\cdot k_y\right)}-1.
	\label{eq:eq06}
\end{equation}

where {$FT^{-1}$} represents inverse Fourier transform, {$\varepsilon$} is a regularization parameter to control the tradeoff between resolution enhancement and noise amplification \cite{choi2017compensationa}, {$H^{*}$} is the complex conjugate of {$H$}. More detailed derivation of Equation (\ref{eq:eq04a}) can be found in Ref.\cite{wang2021quadriwave}.

\begin{figure}[h]%
\centering
\includegraphics[width=0.9\textwidth]{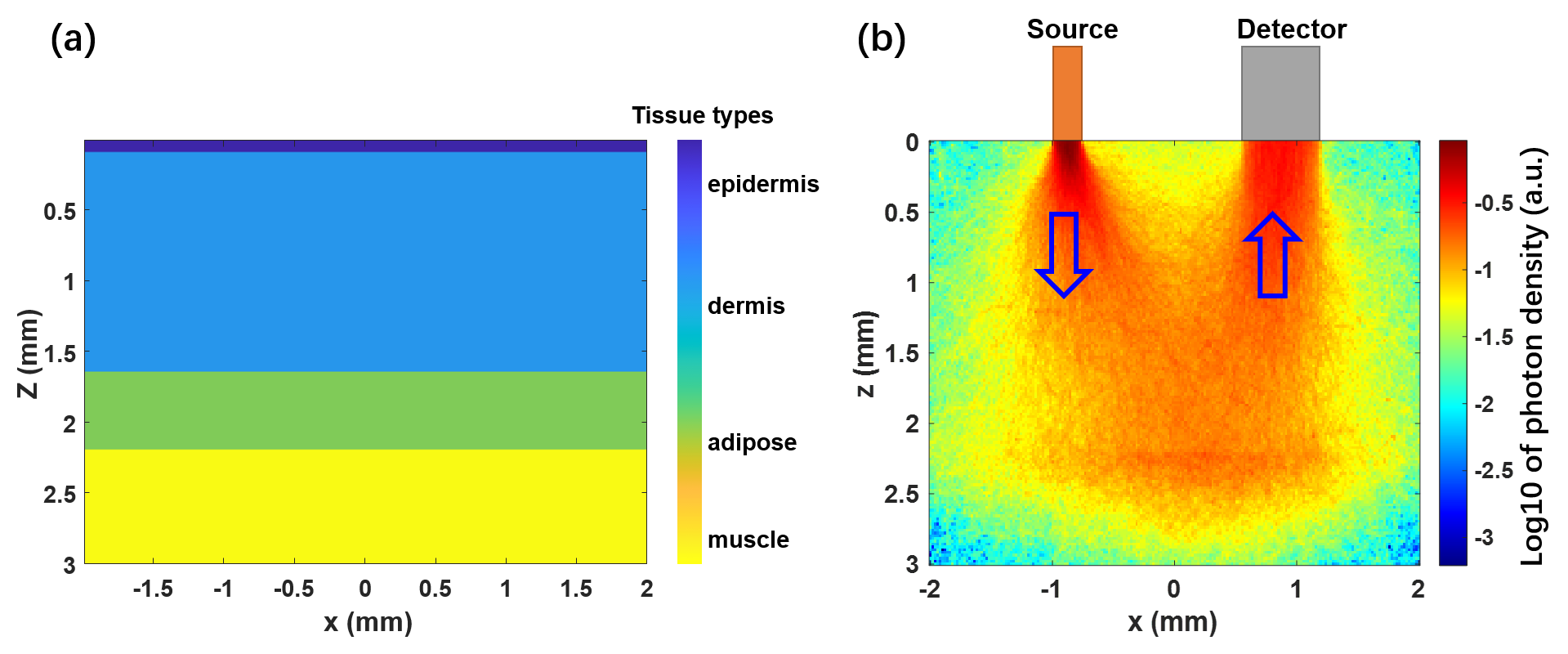}
	\caption{Monte Carlo simulation for photon path through the skin tissue. (a) Illustration of the different skin layers: 0.085 mm epidermis, 1.55 mm dermis, 0.55 mm adipose, and 0.815 mm muscle. (b)The sampling volume is summed along the $y$ axis, which indicates the multiple scattering effect of thick sample turns epi-illumination into virtual transmissive illumination. }
	\label{fig:Monte}
\end{figure}

To understand the meaning of the obtained QPI signal, we perform a Monte Carlo simulation with PyXOpto \cite{Naglic:17,Naglic:15}. Detailed implementation of the simulation is described in Methods. Figure \ref{fig:Monte} (a) shows the simulated object that resembles the anatomy of human skin, using scattering parameters reported in the literature \cite{64354,ding1999analysis}. As shown in Fig. \ref{fig:Monte} (b), the simulation traced the scattering progress of light photons emitted from the illumination fiber and traveling to the detector fiber through the sample. The result of the Monte Carlo simulation suggests that QPI imaging, in this setting, is made possible by the “virtually” trans-illumination field from the deeply scattered light. The imaging depth can be a few hundred microns without severe multiple scattering.

\section{Results}
\subsection{FEM on standard objects}

\begin{figure}[h]%
\centering
\includegraphics[width=0.9\textwidth]{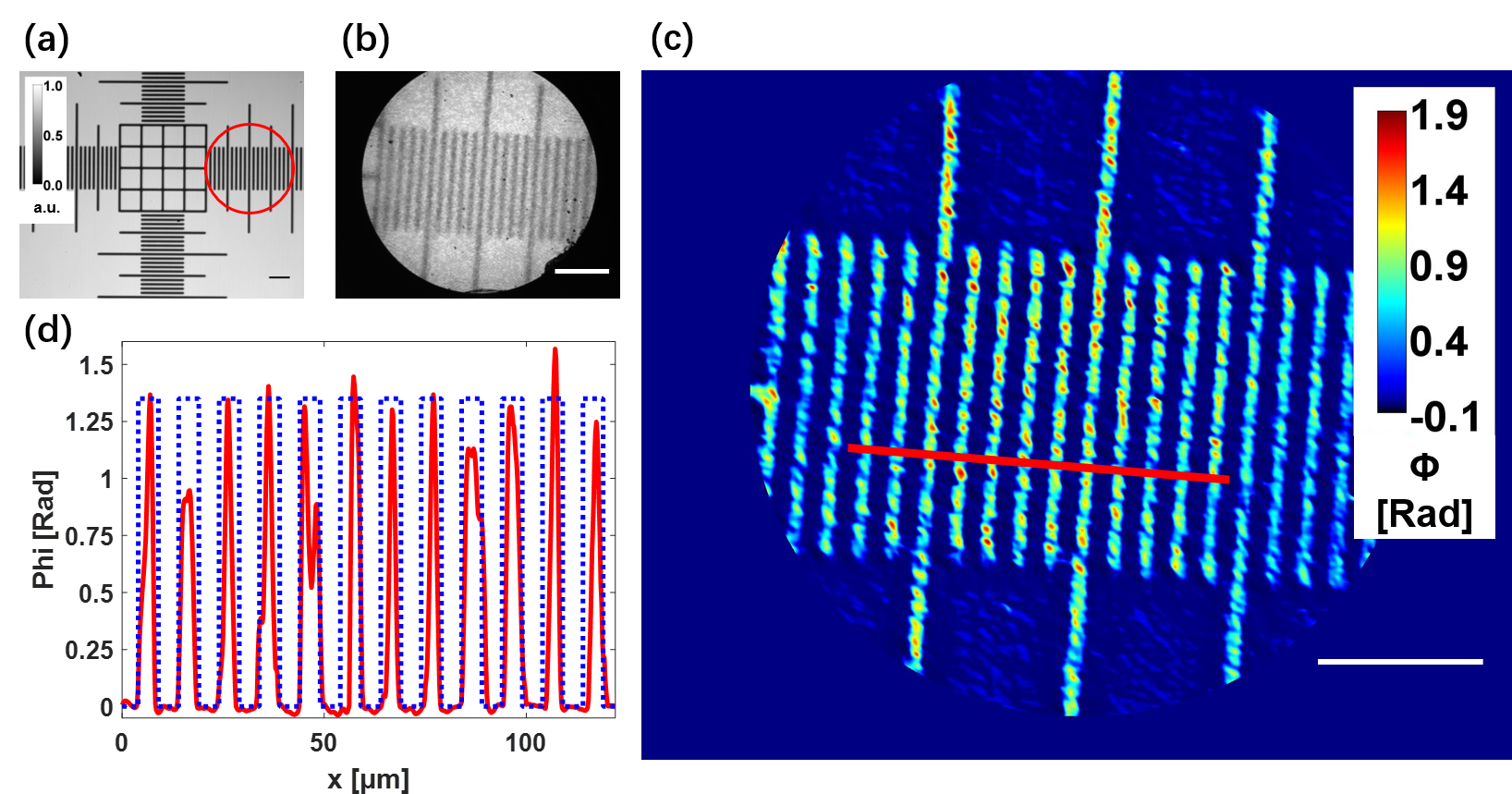}
	\caption{Measurement of 10 {$\mu \mathrm{m}$} scale calibration slide. (a) Microscope image of the slide for reference. (b) Raw endoscopic image. (c) The integrated phase map of the sample. (d) Difference between the simulated and measured phase of the red line in (a). Scale bars 50 {$\mu \mathrm{m}$}.}
	\label{fig:Calibration}
\end{figure}

To evaluate the performance of the FEM system, a calibrated scale was imaged as a testing object. To satisfy the multiple scattering condition simulated in Section 2, a highly scattering substrate (\textit{e.g.}, a stack of paper) is placed underneath the object. Figure \ref{fig:Calibration} (a) shows the object measured under a brightfield microscope, and the red circle indicates the area imaged by FEM. Fig. \ref{fig:Calibration} (b) shows a raw interferogram. The resulting QPI image, reconstructed using Equations (\ref{eq:eq04a})-(\ref{eq:eq06}), is shown in Fig. \ref{fig:Calibration} (c). The spacing between the lines is 10 {$\mu \mathrm{m}$}.  The line profile (red line in Fig. \ref{fig:Calibration} (c)) fits well with the expected phase profile as shown in Fig. \ref{fig:Calibration} (d). More details related to the calculation of the estimated phase can be found in Methods. 

We further investigated the phase sensitivity of the FEM system, which shows the ability to detect the lowest optical path length value from the noise floor. The spatial-temporal noise level, measured to be 16 nm, suggests the FEM is applicable to detect even subtle changes in the refractive index of biological samples with high precision. More detailed experimental results are included in \textit{Supplemental Section 1}.

Additionally, we estimate the resolution of the system using polystyrene beads. Detailed results about this measurement can be found in \textit{Supplemental Section 2}. The lateral and axial resolution of FEM are 1.2 {$\mu \mathrm{m}$} and 8.3 {$\mu \mathrm{m}$}, respectively. This demonstrates the high lateral and axial resolution of FEM in identifying tissue structures.

\subsection{FEM on histological slides}

\begin{figure}[h]%
\centering
\includegraphics[width=0.9\textwidth]{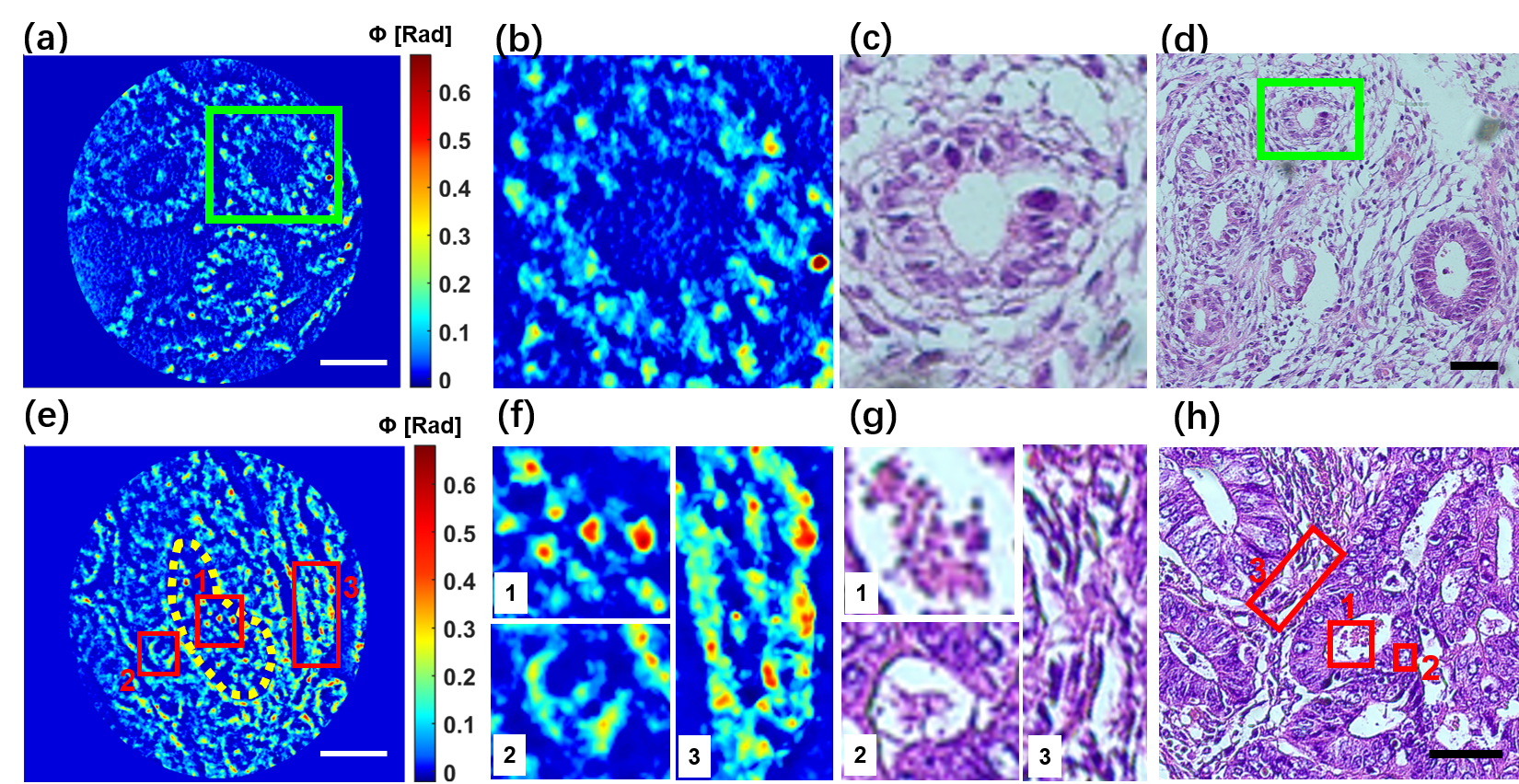}
	\caption{Measurement of hematoxylin and eosin (H{$\&$}E) -stained proliferative endometrium (a-d) and grade 2 endometrial adenocarcinoma (e-h). (a,e) Integrated phase maps of the sample, which points out normal circular gland and tumor irregular-shaped gland. (b,c) Enlarged details of the circular gland. (f,g) Enlarged details of secreta, cell and stroma, labeled with 1-3, respectively. (d,h) Microscopic images under 10× objective. Scale bars 50 {$\mu \mathrm{m}$}.}
	\label{fig:HE}
\end{figure}

To exploit the suitability for biomaterials, we first test FEM on two endometrial slides, one with proliferative endometrium, and the other diagnosed with grade 2 adenocarcinoma. Again, a highly scattering substrate is placed underneath the object to satisfy the multiple scattering condition.  Our specimens, though stained with Hematoxylin and Eosin (H{$\&$}E), can be considered transparent under FEM illumination, given H{$\&$}E's extremely low absorption in NIR-I window  \cite{hong2017nearinfrared}. Figures \ref{fig:HE} (a, b) show FEM images of the proliferative endometrium, where the green box highlights an endometrial gland with regular shape, containing some cell-like materials. The H{$\&$}E images in Fig. \ref{fig:HE} (c, d) validate these characters. Figure \ref{fig:HE} (e) shows the image of the slide with grade 2 endometrial adenocarcinoma, where an irregular-shaped endometrial gland, indicated by the yellow dotted line, shows higher image contrast under FEM, compared to the conventional bright field image shown in Fig. \ref{fig:HE} (h). In Fig. \ref{fig:HE} (f), region 1 shows the secreta in the center of the gland, region 2 shows the tumor cell within the gland, and region 3 is the tumor stroma filled among the glands, which corresponds with  Fig. \ref{fig:HE} (g). The FEM phase images identify normal and cancerous endometrial slides with different pathological structures and gland shapes. 

Besides measuring tissue morphology, FEM can provide further insights into tissue condition by using scattering parameters extracted from QPI images. A demonstration of such tissue analysis is included in \textit{Supplemental Section 3}.

\subsection{FEM on mammalian tissue section \textit{ex vivo}}

\begin{figure}[h]%
\centering
\includegraphics[width=0.9\textwidth]{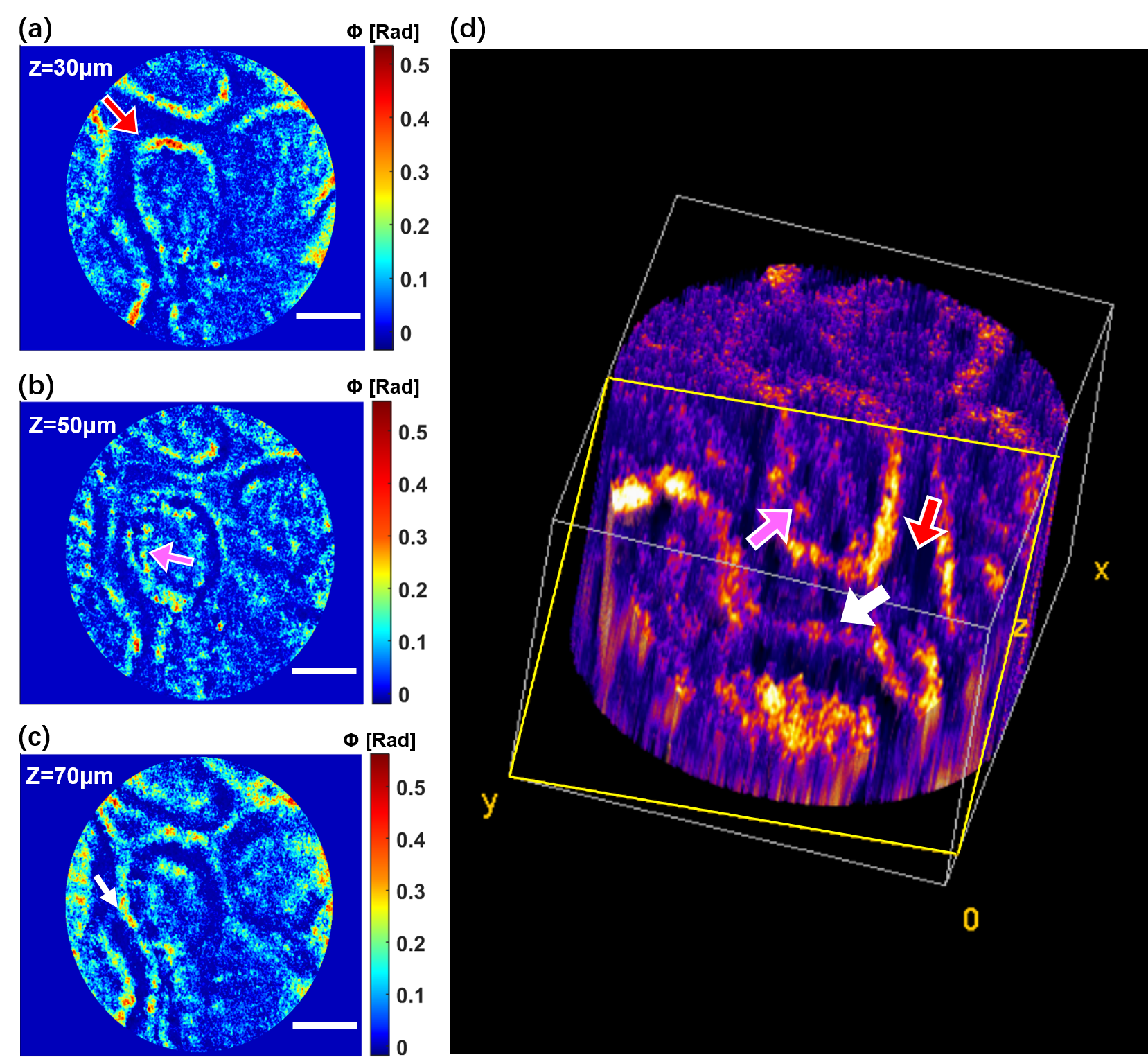}
	\caption{Measurement on the skin of the streaky pork. (a) Phase at stratum spinosum layer. (b) Phase at stratum basal-dermis layer. (c) Phase at dermis layer. (d) The 3D rendering of the phase stack with a size of  250×250×180 {$\mu \mathrm{m}^3$}. The yellow box is a cross-section view. The red arrow denotes canyon in stratum spinosum layer; The pink arrow points the cell in stratum basal-dermis layer; The white arrow shows fibrous components in dermis layer.}
	\label{fig:3D}
\end{figure}

To demonstrate the volumetric imaging performance of FEM, a streaky pork with 10 mm in thickness is used as an example. The procedure of sample preparation can be found in Methods. We performed axial scanning of the sample with a step size of 10 {$\mu \mathrm{m}$} in the $z$-direction. Figures \ref{fig:3D} (a-c), show the transverse FEM phase maps measured at z = 30, 50, and 70 {$\mu \mathrm{m}$}, respectively. In Fig. \ref{fig:3D} (a), the stratum spinosum layer with canyon-like characteristics is clearly visible, indicated by the red arrow. In Fig. \ref{fig:3D} (b), the canyon becomes less distinct and is filled with the stratum basal layer, and the cell-like structure in the dermis is indicated with a pink arrow, representing the stratum basal-dermis layer. Figure \ref{fig:3D} (c) shows fibrous components labeled by the white arrow at the basis of the stratum papillae. These results highlight the capability of FEM to image through thick and turbid samples, revealing textural features at different depths of the layers without the need for stains or fluorescence indicators. Benefiting from annular illumination and coherence gating, the FEM provides good sectioning capability. Figure \ref{fig:3D} (d) presents a 3D rendering of the phase stack, where a cross-section view, indicated by the yellow rectangle, reveals the tomographic details of tissue structured seen in Fig. \ref{fig:3D} (a-c). 

\subsection{FEM on skin tissue \textit{in vivo}}

\begin{figure}[h]%
\centering
\includegraphics[width=0.9\textwidth]{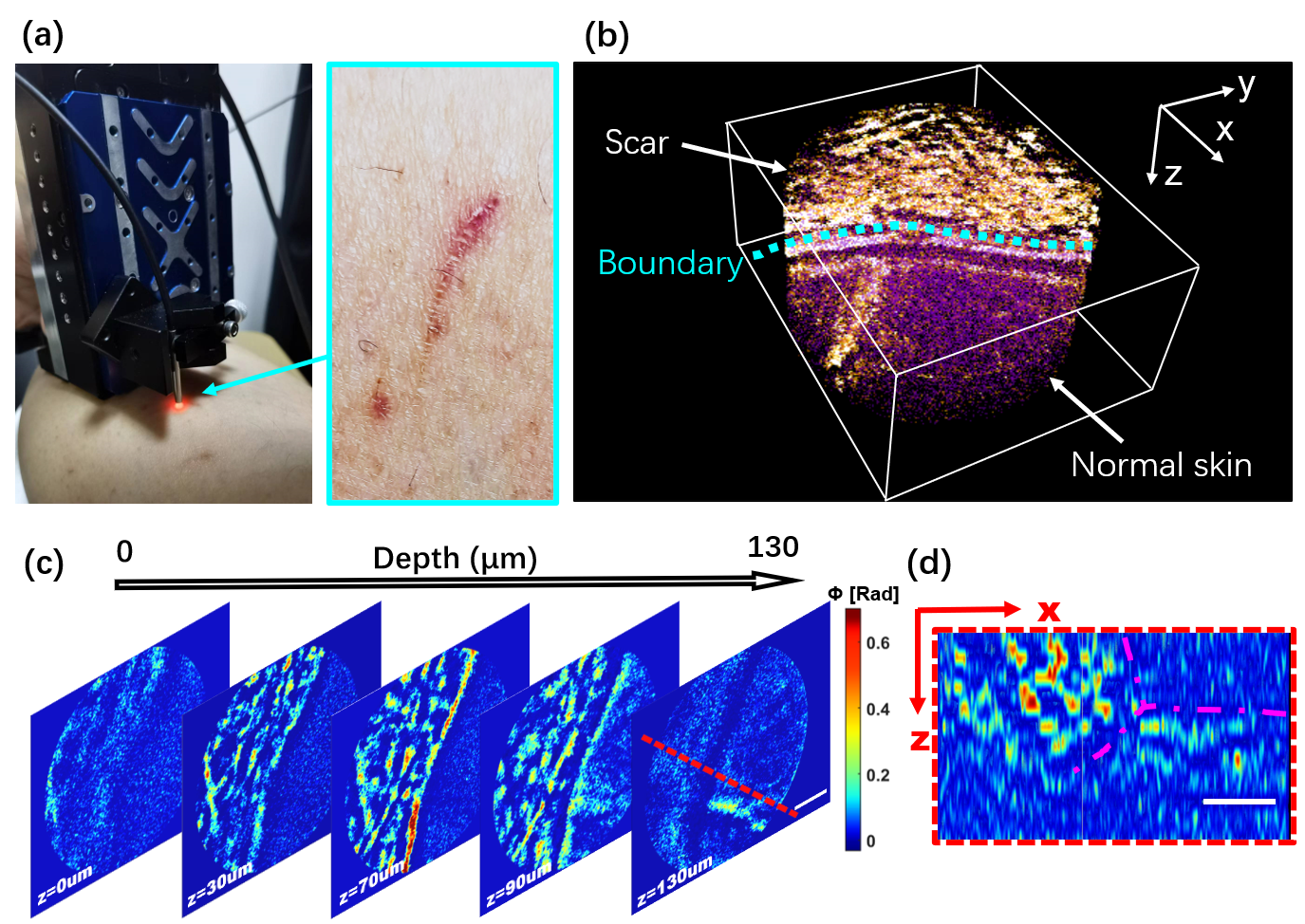}
	\caption{Measurement of the scar. (a) The experimental set up and the scar at 15 days post-wounding (after 2 days of scab removal). (b) 3D renderings of the scar with a size of 250 × 250 × 130 {$\mu \mathrm{m}^3$}. Different focal planes of the scar. (d) The \textit{x-z} plane cross section corresponds to dotted lines in (c). Scale bars, 50 µm. The color bar corresponds to phase values with the unit of radian.}
	\label{fig:vivo}
\end{figure}

Last but not least, to validate the \textit{in vivo} imaging ability, we perform an imaging experiment on human skin. The subject has a skin wound located on the lower limb caused by an animal scratch. Upon incident, the site of the lesion was treated by a medical processional. Figure \ref{fig:vivo} (a) shows a photograph of wounded skin to be measured with FEM, on the 15th day after the treatment. Particularly, we selected a FOV that partially contains scared tissue, as well as unwounded skin. The tissue was scanned axially to a depth of 130 {$\mu \mathrm{m}$}, which is depicted in Fig. \ref{fig:vivo} (c). From skin surface to as deep as 90 {$\mu \mathrm{m}$}, collagen and proliferating fibroblasts form a fibrous union, filling the regenerative layer. At a depth of 130 µm, the scar demonstrates a canyon-like structure that resembles normal skin. Figure \ref{fig:vivo} (d) shows a cross-section image along the red dotted line in Fig. \ref{fig:vivo} (c), clearly illustrating the boundary between the scab fibers and normal skin, marked with a pink dashed line. Figure \ref{fig:vivo} (b) displays a 3D rendering that clearly demonstrates the distinct boundary between the scar and normal skin. The fibrotic tissue is apparent on the upper layer of the scar side (left), while the native layer (right) exhibits uniform tissue. These results highlight the excellent optical sectioning ability of FEM and its effectiveness in illustrating morphological differences between the scar and normal skin. In \textit{Supplemental Section 4}, we also include the measurements on another FOV at a different day. \textit{Supplemental Section 5} discusses \textit{in vivo} imaging of benign nevus.

\section{Summary}

In summary, we developed FEM, a flexible endoscopic QPI system, for quantitative \textit{en face} imaging of turbid biomaterials. The instrument utilizes a small form factor fiber bundle as the endoscope probe and a QPI module located on the proximal end for imaging, whereas quantitate images are reconstructed via postprocessing. The FEM directly measures the phase of light transmitted through the sample, and it does not require intensive modeling of the system transfer function and inverse reconstruction. Due to the slow response of the existing SLM (100 ms switching time), the imaging speed of the current setup is slow. However, we believe this limitation can be overcome by employing a more state-of-art device (such as ferroelectric liquid crystal on silicon devices from Forth Dimension Displays \cite{4dd}, or a metasurface to perform instantaneous wavefront modulation\cite{kwon2020singleshot,zhong2020quantum}.

A series of experiments have been conducted to demonstrate the potential of FEM for applications in biomedicine field. By measuring calibrated scale and microspheres, the proposed instrument demonstrates high spatial-temporal stability with cellular level resolution in both transverse and axial directions. In addition, FEM was used to measure benign and diseased histological slides, and our QPI images show high quality outcome in identifying pathological relevant tissue structures, \textit{e.g.}, benign and malignant endometrial glands. Furthermore, built upon its intrinsic sectioning capability, our experiments on \textit{ex vivo} and \textit{in vivo} mammalian skin tissue highlight FEM capabilities in distinguishing epidermis and dermis layers with characteristics such as canyons and fibrous tissue at different depths. Such label-free imaging modality could potentially expedite preoperative diagnosis by reducing the time in staining and tagging as used in conventional methods. Although the experiments were conducted in open space, the flexible probe with a small form factor (2.5 mm used in this study) suggests FEM can navigate through regions with limited access, \textit{e.g.}, bronchi, with minimal invasiveness. With the quantitative information, high sensitivity, and label-free imaging capabilities, we believe FEM can greatly benefit the biomedical community.

\section{Methods}

\subsection{FEM System Configuration}
The optical setup is shown in Fig. \ref{fig:setup}. The LED for illumination has a central wavelength of 780 nm and a full-width half-maximum of 30 nm, which is coupled into the light guide and illumination fibers. The illumination fibers (Sumita, SOG-120C) with the divergent angle of 120° surround the fiber bundle to create the ring illumination. To ensure sufficient illumination area while minimizing the probe tip size, the illumination fibers have a higher numerical aperture (NA) and form a smaller ring width compared to the fiber bundle. For image collection, a gradient refractive index (GRIN) lens (GRINTECH, NEM-100-25-10-860-DS, NA=0.5, 2.6×) with an outer diameter of 1 mm works as the micro-objective. The working distance of the GRIN lens is about 180 {$\mu \mathrm{m}$} in the air and 250 {$\mu \mathrm{m}$} in the water. The collected image is then transferred by a 1-m-long fiber bundle (Fujikura, FIGH-30-650S) with 30,000 cores. The average core diameter is 3 {$\mu \mathrm{m}$}, and core-to- core spacing is about 3.5 {$\mu \mathrm{m}$}. The outer diameter of the fiber bundle is 750 {$\mu \mathrm{m}$}, and the imaging diameter is 650 {$\mu \mathrm{m}$}. The fiber bundle tip was fixed on a linear stage (Physik Instrumente, V-528) for axial scanning. The objective (Olympus UPLXAPO, NA=0.4, 10×) accompanied with the tube lens collect images from the fiber bundle for further phase modulation. As the fiber cores act as sampling pattern with a period of 3.5 {$\mu \mathrm{m}$}, the 10× objective has a magnification of 10, which meets the Nyquist sampling condition for the image field with an ideal resolution of 1.2 {$\mu \mathrm{m}$}. To modulate the phase of the field, an amplitude type SLM (UPOLabs, HDSLM80RA, 60 Hz) is used to display chessboard grating. The grating has a period of 2400 pixels on the SLM. A CMOS (Basler, acA1300-60gmNIR, 60 Hz) records the interferograms of the differentiated wavefront. The phase reconstruction algorithm is real-time. The phase imaging speed of the system is 2 Hz, mostly due to the low SLM stabilization time of 100 ms.

\subsection{Monte Carlo Simulation}
We utilized PyXOpto, an opensource software, to perform Monte Carlo simulation. To simplify the simulation, we only model the left half of the probe cross-section, and replace the annular illumination fiber bundle with a single illumination fiber, and the entire imaging bundle with a single imaging fiber. The detector fiber bundle had a diameter of 650 {$\mu \mathrm{m}$}, NA of 0.5, and a refractive index of 1.45, while the illumination fiber bundle had a diameter of 250 {$\mu \mathrm{m}$}, NA of 0.9, and a refractive index of 1.45. The distance between the two fiber cores was set to 875 {$\mu \mathrm{m}$}. To visualize the path of photon packets through the sample, we used a tool called sampling volume, which summed the photon packet weights along the $y$-axis.

\subsection{Calibrated scale ideal phase and shear distance estimation}

We calculate the ideal phase of the calibrated scale with {$\varphi_0\left(x,y\right)=2\pi\left(n-n_0\right)d/\lambda$}, where $\lambda$ = 780 nm is the central wavelength, $n$ = 1.51 is the refractive index of the glass, $n_0$ = 1 is the refractive index of the air, $d$ = 340 nm ± 20 nm is the height of the scale. In the phase calculation process, we use the shear distance value $x_0$ = $y_0$ = 373 nm. The shear distance estimation process can refer to Ref. \cite{nguyen2017gradient}.

\subsection{Sample Preparation and Imaging}

\subsubsection{Streaky pork preparation}
The streaky pork was washed with phosphate buffered saline (PBS, Biosharp) and soaked in 4 $\%$ paraformaldehyde (Biosharp) for one day at room temperature. Then, the fixed sample was washed with PBS and placed whole on a glass slide for imaging. The axial scanning step size is 10 {$\mu \mathrm{m}$}.

\subsubsection{\textit{In vivo} imaging of scar}
The patient had a cat scratch on her lower leg. The wound was washed with 20 $\%$ soapy water for 15 minutes within 1 hour of injury and then sterilized with Iodophor. The scar was imaged at 15 days post-wounding (2 days post scab removal). To minimize bulk motion artifacts introduced by the awake patient, we placed the linear stage holding the fiber probe on the skin surface. The camera exposure time is 10 ms. The linear stage had a waiting time of 5 ms and a default scanning speed of 5 mm/s.

\backmatter
\bmhead{List of Supplemental Materials}
1. Supplementary Information

\bmhead{Data availability}
The data that support the findings of this study are available from the corresponding author upon reasonable request.

\bmhead{Code availability}
The MATLAB code used for phase reconstruction is available from the corresponding authors upon reasonable request.

\bmhead{Acknowledgements}
This work is supported by National Key R{$\&$}D Plan Strategic Cooperation Key Special Project (2023YFE0203200), National Natural Science Foundation of China (52275533), Key R{$\&$}D Plan of Hubei Province (2023BCB085, 2021BAA056), Innovation Project of Optics Valley Laboratory (OVL2023PY001) and Shenzhen Technical Project (JCY J20210324141814038). We need to mention that Chenfei Hu has no relationship with financial sponsorship. We would like to thank Dr. Yanli Ouyang from Fifth Hospital in Wuhan for providing histological slices. We would like to express great appreciation to the late Professor Gabriel Popescu for his kind constructive advising and discussion.  

\bmhead{Author contributions}
J.W. proposed the idea, constructed the instrument, wrote the algorithm, performed imaging, and analyzed the data. J.W. and C.H. wrote the paper, processed data, and draw figures. W.Y. and Y.J gave support on instrument setup. Y.Z. provided biospecimen analysis. X.L., X.J., and W.L. supervised the work.

\section*{Declarations}
The authors declare no conflicts of interest.


\bibliography{library_abbreviated}

\end{document}